\documentclass{article}
\usepackage[utf8]{inputenc}
\usepackage[a4paper, margin=1in]{geometry}
\usepackage{graphicx}
\usepackage{multirow}
\usepackage[table,xcdraw]{xcolor}
\usepackage[round]{natbib}
\usepackage[colorlinks=true, allcolors=blue,hypertexnames=false]{hyperref}
\usepackage[font=small,labelfont=bf]{caption}
\usepackage[section]{placeins} 

\usepackage{mathptmx}
\usepackage{babel}
\usepackage{amsmath}
\usepackage{amsfonts}
\usepackage{authblk}

\usepackage{booktabs}

\title{Benchmarking Retinal Blood Vessel Segmentation Models for Cross-Dataset and Cross-Disease Generalization}

\author[1,2]{Jeremiah Fadugba}
\author[3]{Patrick K{\"o}hler}
\author[3,4]{Lisa Koch}
\author[5]{Petru Manescu}
\author[3,6]{Philipp Berens}

\affil[1]{University of Ibadan, Nigeria.}
\affil[2]{African Institute for Mathematical Sciences, Rwanda.}
\affil[3]{Hertie Institute for AI in Brain Health, University of T\"ubingen, Germany.}
\affil[4]{Department of Diabetes, Endocrinology, Nutritional Medicine and Metabolism UDEM, Inselspital, Bern University Hospital, University of Bern, Switzerland.}
\affil[5]{University College London, London.}
\affil[6]{Tübingen AI Center, T\"ubingen, Germany.}

\date{}

\begin{document}

\maketitle

\begin{abstract}
\noindent
\textbf{Purpose:}
Retinal blood vessel segmentation can extract clinically relevant information from fundus images. As manual tracing is cumbersome, algorithms based on Convolution Neural Networks have been developed. Such studies have used small publicly available datasets for training and measuring performance, running the risk of overfitting. Here, we provide a rigorous benchmark for various architectural and training choices commonly used in the literature on the largest dataset published to date. \\
\textbf{Methods:} 
We train and evaluate five published models on the publicly available FIVES fundus image dataset, which exceeds previous ones in size and quality and which contains also images from common ophthalmological conditions (diabetic retinopathy, age-related macular degeneration, glaucoma). We compare the performance of different model architectures across different loss functions, levels of image qualitiy and ophthalmological conditions and assess their ability to perform well in the face of disease-induced domain shifts.\\
\textbf{Results:} Given sufficient training data, basic architectures such as U-Net perform just as well as more advanced ones, and transfer across disease-induced domain shifts typically works well for most architectures. However, we find that image quality is a key factor determining segmentation outcomes. \\
\textbf{Conclusions:} When optimizing for segmentation performance, investing into a well curated dataset to train a standard architecture yields better results than tuning a sophisticated architecture on a smaller dataset or one with lower image quality.\\
\textbf{Translational Relevance:} We distilled the utility of architectural advances in terms of their clinical relevance therefore providing practical guidance for model choices depending on the circumstances of the clinical setting.

\end{abstract}

\section{Introduction}

Retinal fundus imaging is the only noninvasive instrument that allows us to identify geometric characteristics of a deeper microvascular system such as vessel diameters or branching angles. These techniques are not only employed in the diagnosis of retinal diseases such as glaucoma, diabetic retinopathy (DR), and age-related macular degeneration (AMD) \citep{10.1167/tvst.11.7.12}, but also in the diagnosis of microvascular conditions such as hypertension and atherosclerosis \citep{fathi2013automatic, kanski2011clinical}. While hypertension, for instance, can lead to remodeling of blood vessels, diseases like diabetes can cause new vessels to appear. Therefore, segmenting the vessels from the fundus images, that is, delineating their boundaries from the background, is an important initial step in several diagnostic procedures.

Manual segmentation of retinal vessels is a time-consuming process, requiring approximately three to five hours per image \citep{fives}. It has been shown that machine learning algorithms can be used to effectively solve numerous medical segmentation tasks effectively \citep{salpea2022medical, yao2024cnn, Ma2024-jo}, including retinal vessel segmentation \citep{hegde2023systematic}. Recently, deep learning approaches have demonstrated the capacity to achieve human-level performance in this task \citep{LIN2023102929}.
The majority of these approaches are based on the UNet model \citep{unet}. Some of these approaches include architectural modifications tailored to the specific task of segmenting retinal blood vessels \citep{FR-Unet, little-wnet, saunet}, which  requires the ability to detect relatively thin objects with a high degree of connectivity. While all models share the ability to encode the image efficiently into a more compact representation, they differ in the inductive biases according to which they are designed. For example, FR-Unet \citep{FR-Unet} aims to retain the full resolution of the image, whereas MA-Net \citep{manet} places greater emphasis on the encoded dependencies between local and global image features.

Despite the plethora of different approaches, there is currently no direct comparison that highlights the respective strengths and weaknesses of the different models for a wide range of clinical settings. Moreover, existing work on retinal vessel segmentation has been evaluated on few and relatively small publicly available datasets \citep{DeepVessel, Zhao-et-al, zhou2018unet, Wangetal(MVP)}. This is in part due to the absence of large publicly available retinal image datasets with annotated blood vessels. In addition to their small sample size, the most commonly studied datasets, DRIVE \citep{drive_data} and CHASEDB1 \citep{chase_data}, vary strongly in their image quality and labeling protocols \citep{little-wnet}. Furthermore, they contain few images of patients with retinal disease, such that it is currently unclear how the available models generalize to real patient populations with differing characteristics in terms of retinal diseases, image quality, and varying dataset sources. 

Given the wide range of choices for architectures, loss functions, or training protocols, it is also unclear which of these factors actually matter when a large dataset is used for training. 

In this paper, we present a comprehensive survey and benchmark of the current state of the art in retinal blood vessel segmentation on fundus images. We investigate the most commonly used configurations of model architectures and loss functions for model training and evaluate the segmentation performance on three datasets, including the largest publicly available dataset \citep{fives} as well as previously well-established smaller benchmark datasets \citep{drive_data,chase_data}. We explicitly test whether the different models are robust to the inherent variability of fundus images due to different diseases or changes in image quality.

\section{Methods}

\subsection{Data}
\begin{figure}
    \centering
    \includegraphics[width=0.9\linewidth]{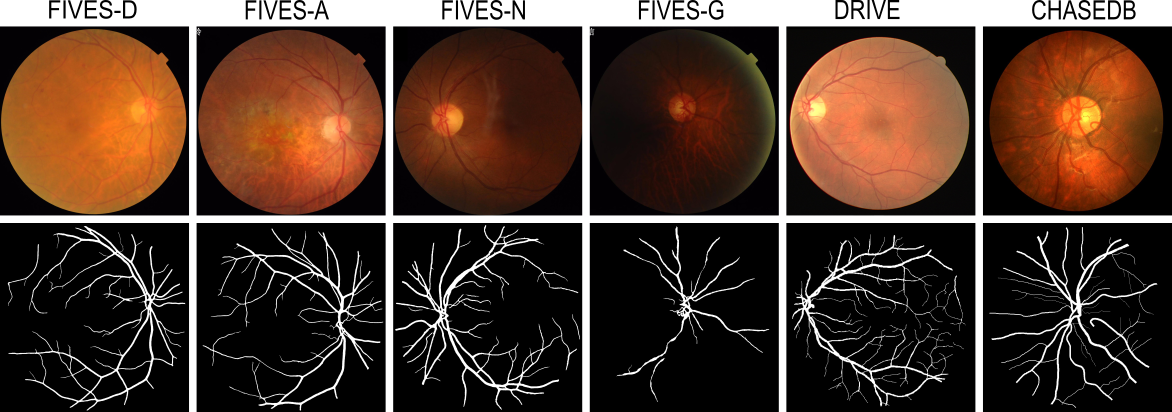}
    \captionof{figure}{Example images and manual segmentations for FIVES (including an example from each subgroup), DRIVE and CHASEDB1.}
    \label{fig:dataset_masks}
\end{figure}

We developed and evaluated our models on three publicly available retinal fundus datasets for vessel segmentation (Fig. \ref{fig:dataset_masks}, Table \ref{tab:dataset_summary}). We used the largest high-quality dataset to date for Fundus Image Vessel Segmentation (FIVES) \citep{fives}, which has so far not been used to analyze segmentation models. It consists of 800 retinal images from 573 patients. Retinal vessels were manually annotated in a consensus procedure involving  three senior ophthalmologists and 24 medical residents. The dataset includes disease labels for AMD (``A'', 200 images), DR (``D'', 200), Glaucoma (``G'', 200) and Healthy (``N'', 200), as well as manually assessed image quality grades. Image quality was attributed to either illumination and color (\textcolor{blue}{$644$ }images), blur (\textcolor{blue}{$669$} images) or low contrast (\textcolor{blue}{ $765$} images). 
 
In addition, we used well-known fundus datasets for further evaluation: The DRIVE dataset \citep{drive_data} consists of 40 fundus images obtained from a diabetic retinopathy screening program, of which seven images (not further specified) contained pathologies. Finally, the CHASEDB1 dataset \citep{chase_data} consists of 28 fundus images of the left and right eyes of 14 children. 

\begin{table}
    \centering
    \caption{Summary of publicly available retinal vessel segmentation datasets used in this study including the prevalence of images with diabetic retinopathy (``D''), age-related macular degeneration (``A''),  glaucoma (``G''), as well as healthy (``N'') images.}
    \label{tab:dataset_summary}
    \begin{tabular}{ccccccc}
        \toprule
        Dataset & Year & Images & Resolution & Disease prevalence & Annotators & Train/Test split\\
        \midrule
        FIVES & 2021 &  800 & $2048\times 2048 $ & 200 N, 200 A, 200 D, 200 G & Group & 600/200 \\
        DRIVE & 2004  & 40 & $768 \times 584 $ &  33 N, 7 D & 3 & 20/20 \\
        CHASEDB1 & 2011 & 28 & $990 \times 960$  & 28 N & 2 & 20/8\\
        \bottomrule
    \end{tabular}

\end{table}

All images were pre-processed using Contrast Limited Histogram Equalization (CLAHE) \citep{pizer1987adaptive} with a clip limit of 2 and a grid size of $8 \times 8$. The images and segmentation masks were resampled to $512\times512$ pixels. Finally, the datasets were divided into training and test folds according to the ``official'' splits. 

\subsection{Segmentation model architecture}

We performed a comprehensive literature search and selected 22 articles on retinal vessel segmentation published between 2016 and 2022 (see Table\,\ref{table-lit-review}). Based on their reported segmentation performance and the availability of code, we included the standard UNet segmentation model and four variants in our benchmark. In the following, we briefly outline the respective model architectures. 

\begin{table}
\vskip 0.15in
\begin{center}
\begin{small}
\begin{sc}
\caption{Self-reported performance of some existing vessel segmentation methods from our literature survey. We report the loss used for training, reported Dice (DSC), Area under the Curve (AUC), and Matthew's Correlation Coefficient (MCC) metrics for models evaluated on DRIVE and CHASEDB1, where available. We also show which models reported their  performance at native resolution  and omit others that either did not report  which approach they used or which used a different evaluation approach. The best results are shown in bold.}
\label{table-lit-review}
\centerline{
\begin{tabular}{lccccccccc@{}}
\hline
\multicolumn{3}{c}{}       & \multicolumn{3}{c}{DRIVE} & \multicolumn{3}{c}{CHASE-DB1} \\ 
\cmidrule{4-6}\cmidrule{7-9}

\multicolumn{1}{c}{Method}& Resolution & Loss & DSC      & AUC    & MCC    & DSC       & AUC      & MCC     \\ \hline
\citep{DeepVessel}        & -  & - & 78.75   & 94.04  & -      & 75.49    & 94.82    & -       \\
\citep{LADOS}             & -  & - & -       & 96.36  & -      & -        & 96.06    & -       \\
\citep{Orlandoetal}       & -  & - & 78.57   & 95.07  & 75.56  & 73.32    & 95.24    & 70.46   \\
\citep{linGuetal}         & -  & - & 78.86   & -      & 75.89  & 72.02    & -        & 69.08   \\
\citep{MS-NFN}            & -  & - & -       & 98.07  & -      & -        & 98.25    & -       \\
\citep{yanetal}           & -  & - & 81.83   & 97.52  & -      & -        & 97.81    & -       \\
\citep{laddernet}         & -  & BCE & 82.02   & 97.93  & -      & 80.31    & 98.39    & -       \\
 \citep{Wangetal(MVP)}    & -  & - & 81.44   & -      & 78.94  & 78.63    & -        & 76.55   \\
\citep{DEU-Net}(DEU-Net)   & -  & DiceBce & 82.70   & 97.72  & -      & 80.37    & 98.12    & -       \\
\citep{DVAE-Refiner}       & -  & BCE & -       & 97.90  & -      & -        & 98.20    & -       \\
\citep{fu-et-al}           & -  & - & 80.48   & 97.19  & -      & -        & -        & -       \\
\citep{Wang-et-al}          & -  & - & 80.93   & -      & 78.51  & 78.09    & -        & 75.91   \\
\citep{CE-Net}              & -   & DICE & -       & 97.79  & -      & -        & -        & -       \\
\citep{R-sGAN}(R-sGAN)      & -  & - & 78.82   & -      & -      & -        & -        & -       \\
\citep{M2U-Net}(M2U-Net)    & -  & DiceBce & 80.91   & 97.14  & -      & 80.06    & 97.03    & -       \\
\citep{shin-et-al}          & -  & - & 82.63   & 98.01  & -      & 80.34    & 98.30    & -       \\
\citep{iternet}(iter-Net)   & Native  & - & 82.05   & 98.16  & -  & 80.73    & 98.51    & -       \\
\citep{saunet} (SA-UNet)    & Native  & BCE & 82.63   & 98.64  & \textbf{80.97}  & 81.53 & 99.05 & \textbf{80.33} \\
\citep{SGL}                  & - & - & 83.16   & 98.86  & -     & 82.71    & 99.20     & -       \\
\citep{RV-GAN} (RV-GAN)      & - & - & \textbf{86.90}   & 98.87  & -      & \textbf{89.57}    & 99.14    & -       \\
\citep{Galdran2022} (W-Net)  & Native & BCE & 82.79   & 98.10   & 80.24  & 81.69    & 98.47    & 79.74   \\
\citep{FR-Unet}(FR-UNet)     & Native & BCE & 83.16   & \textbf{98.89}  & -      & 81.51    & \textbf{99.20} & -  

\end{tabular}
}
\end{sc}
\end{small}
\end{center}
\vskip -0.1in

\end{table}

\paragraph{UNet} The standard UNet \citep{unet} is the most common baseline algorithm for segmentation tasks including vessel segmentation. Its architecture is characterized by a U-shaped network structure consisting of an encoder and a decoder. The encoder is responsible for capturing the features of the input image.
The decoder, on the other hand, takes the learned features from the encoder and gradually upsamples them to the original input image size. The purpose of the decoder is to map the result of the encoder onto a pixel-wise segmentation map that has the same dimensions as the input image. One of the innovations of the UNet is the use of skip connections. These  allow the decoder to access feature maps from the encoder at multiple resolutions. In this way, the UNet combines both high-level and low-level features during the decoding process, helping to capture fine-grained details in the segmentation output. The final output of the UNet is a pixel-wise segmentation map where each pixel is assigned a probability of belonging to the foreground.

\paragraph{FR-UNet}
The FR-UNet architecture \citep{FR-Unet} is the current state-of-the-art model for vessel segmentation. It is based on the idea of preserving the full resolution of the image during training. This is achieved by expanding the network horizontally and vertically through a multi-resolution convolution mechanism.

\paragraph{MA-Net}
The multi-attention UNet (MA-Net) \citep{manet} was initially developed for liver and tumor segmentation and introduces a self-attention mechanism to adaptively integrate local features with their global dependencies. It also aims to exploit the full resolution of the image to capture rich information during end-to-end training but, unlike FR-UNet, it can capture rich contextual dependencies based on the attention mechanism, using two blocks; a Position-wise Attention Block (PAB), which captures the spatial dependencies between pixels in a global view, and a Multi-scale Fusion Attention Block (MFAB), which captures the channel dependencies between arbitrary feature maps through multi-scale semantic feature fusion. 

\paragraph{SA-UNet}
The Spatial Attention UNet (SA-UNet)  \citep{saunet} modifies the standard UNet architecture by introducing an attention module at the spatial dimension of a UNet architecture. This attention module learns the spatial relationship between features from the encoder stage. The spatial attention is derived by applying max-pooling and average-pooling operations along the channel axis and is then concatenated to produce an "efficient feature" map. The features from this stage are then passed to the decoder.

\paragraph{W-Net}
W-Net \citep{little-wnet} is a cascaded extension of the UNet architecture. It involves the consecutive use of the UNet architecture so that an input image is passed through a standard UNet and the resulting output is concatenated with the input image, and passed again through a second UNet to produce the final output. This cascading approach enhances prediction performance but suffers from doubling the number of parameters in the network, thereby increasing the computational requirements. W-Net avoids this by ensuring that the base UNet model is shallow by reducing the number of layers in the network.

\subsection{Loss functions used for training} 

For retinal vessel segmentation, various loss functions have been used, including the Binary Cross-Entropy (BCE), the SoftDice loss, the DiceBCE loss and the centerline-Dice score. 

The \textit{Binary Cross-Entropy} (BCE) loss regards segmentation  as a pixel-wise binary classification problem. For a ground truth segmentation $y$ and predicted probabilities $\hat y$, the BCE loss is defined as 
\begin{equation}
    \mathcal{L}_{\textrm{BCE}}(y, \hat{y}) := - \frac{1}{N}\sum_{i=1}^{N} y_i \cdot \log(\hat{y}_i) + (1 - y_i) \cdot \log(1 - \hat{y_i}). 
\end{equation}
Here, $N$ is the total number of pixels in $y$ and the $i$-th pixel is indexed by $i$ and we assume \mbox{$y_i \in \{0, 1\}$}. For foreground pixels $(y_i = 1$), only the predicted log-probability $\log(\hat{y}_i)$ of belonging to the foreground contribute to the loss, while for background pixels $(y_i = 0)$, the contribution is the log-probability of belonging to the background, $\log(1 - \hat{y}_i)$.

The \textit{SoftDice} loss function \citep{7785132} reformulates the Dice Similarity Score, a popular evaluation metric for image segmentation, to make it differentiable with respect to the predicted probabilities, such that it can be used as an optimization criterion.
It is defined as
\begin{equation}
    \mathcal{L}_\textrm{SoftDice}(y, \hat{y}) := 1 - \frac{2 \sum_{i=1}^{N} y_i \cdot \hat{y}_i + \epsilon}{\sum_{i=1}^{N} y_i + \sum_{i=1}^{N} \hat{y}_i + \epsilon},
\end{equation}
where \mbox{$\epsilon > 0$} is a smoothing factor to prevent division by zero.

The \textit{DiceBCE} loss function is a combination of both the SoftDice and the Binary Cross-Entropy. Previous work has shown that an optimal combination of these two loss functions improves performance \citep{liu2022hidden, MA2021102035, galdran2022optimal}. The \textit{DiceBCE} is calculated as follows:
\begin{equation}  \mathcal{L}_\textrm{DiceBCE}(y, \hat{y}) = \alpha \cdot \mathcal{L}_{\textrm{BCE}}(y, \hat{y}) + \beta \cdot \mathcal{L}_\textrm{SoftDice}(y, \hat{y}) ~,
\end{equation}
where \( \alpha \) and \( \beta \) are weighting factors that balance the contribution of each loss.

In addition, we use a loss function based on the centerline Dice (\textit{clDice}) \citep{cl_dice}, a similarity measure that is calculated from the intersection of the morphological skeletons $S_{\hat{Y}}, S_Y$  of the predicted and ground truth segmentations $\hat{Y}, Y$. The rationale behind incorporating the vessel structure into the loss function is to preserve tiny vessels, which contain only a small area of important structural information and therefore contribute only little to BCE or DSC-based losses if omitted or annotated wrongly, and to enforce the connectivity of the vessels, which is not considered by Dice scores.
As proposed in \citep{cl_dice}, we first compute the the fraction of $S_Y$ that lies in $\hat{Y}$, given as $S_Y2\hat{Y} := \frac{\lvert S_Y\cap \hat{Y} \rvert }{\lvert S_Y \rvert}$ and consequently, the fraction of $S_{\hat{Y}}$ that lies in $Y$, given as $S_{\hat{Y}}2Y := \frac{\lvert S_{\hat{Y}}\cap Y \rvert }{\lvert S_{\hat{Y}} \rvert}$. The \textit{clDice} is then defined as
\begin{equation}
    \textrm{clDice}(Y, \hat{Y}) := 2\frac{S_{\hat{Y}}2Y \cdot S_Y2\hat{Y}}{S_{\hat{Y}}2Y + S_Y2\hat{Y}} ~.
\end{equation}

The clDice can be made differentiable using an iterative soft skeletonization approach \citep{cl_dice} which can be applied as a proxy for morphological erosion and dilation. The resulting \textit{soft-clDice} can be incorporated as an auxiliary loss term to enforce morphological similarity:

\begin{equation}
    \mathcal{L}(y, \hat{y}) := (1- \alpha )\mathcal{L}_{\textrm{SoftDice}}(y, \hat{y}) + \alpha (1 - \textrm{softclDice} (y,\hat{y}))
\end{equation}

\subsection{Training details}
We used data augmentation techniques, including random horizontal and vertical flips with a fixed random rotation during training. 
All model training was carried out using PyTorch on a single NVIDIA-GeForce RTX 2080ti GPU. The models were trained for 70 epochs with a batch size of 4. We employed the Adam optimizer, incorporating a weight decay factor of $1 \times 10^{-5}$, an initial learning rate set to $1 \times 10^{-4}$, and a cosine annealing strategy for learning rate scheduling. 

\subsection{Measures used for evaluation}

To evaluate the performance of the models, we used the Dice Similarity Coefficient (DSC) and the Matthews Correlation Coefficient (MCC). The evaluation was performed on the official test set of each dataset.

\subsection{Code and data availability}
All implemented models, analysis and visualization code is available on GitHub \footnote{\url{https://github.com/berenslab/Retinal-Vessel-Segmentation-Benchmark}}. The datasets FIVES\footnote{\url{https://shorturl.at/WQyGV}}, DRIVE\footnote{\url{https://drive.grand-challenge.org/}}, and CHASEDB1\footnote{\url{https://blogs.kingston.ac.uk/retinal/chasedb1/}} used in this study are officially available at the referenced repositories.

\section{Results}

We first analyzed the performance of five state-of-the-art models trained with different loss functions in a classical in-domain setting using the FIVES dataset \citep{fives}, which is the largest dataset of fully annotated fundus images available. We then investigated the generalization capabilities of the different models and studied their performance in a cross-dataset setting. Finally, we compared how the performance of these models varied under different ophthalmological conditions and for different image quality levels.

\subsection{Choice of architecture and training loss for vessel segmentation}

We trained five prominent segmentation model architectures including a standard UNet on the FIVES training split using four different loss functions (see Methods). We then evaluated their performance on the test split of the same dataset using the Dice Coefficient (DSC) and the Mathews Correlation Coefficient (MCC). 

Regardless of the loss function or the evaluation measure, the UNet, FR-UNet and MA-Net performed similarly and clearly better than the SA-UNet and W-Net (Table~\ref{tab:model_results_various_loss}). For example, the top three models achieved a DSC of about 0.9 regardless of the loss function, while the other two models achieved a DSC of about 0.85. The loss function had a comparably minor influence on the final performance, with DiceBCE and clDice leading to slightly better performing models.

The performance of the top three models was thus close to the reported inter-rater consistency among junior graders at a DSC of 0.92 but slightly lower than the inter-rater consistency between a senior grader and several junior graders at a DSC of 0.96 \citep{fives}. Thus, the best models achieved near-human performance when trained on a large dataset, but surprisingly, none of the architectural variants were able to meaningfully improve upon the baseline UNet architecture.

\begin{table}[htbp]
\caption{Performance of our implementation of the studied segmentation models on the FIVES dataset across various loss functions.}
\centering
\setlength{\tabcolsep}{0pt} 
\begin{tabular*}{\textwidth}{@{\extracolsep{\fill}\quad}lcccc@{\hspace{40pt}}ccccc}
\toprule
& \multicolumn{4}{c}{DSC} 
& \multicolumn{4}{c}{MCC}\\

& BCE & Dice & DiceBCE & clDice 
& BCE & Dice & DiceBCE & clDice\\
\midrule
\multicolumn{1}{l}{\itshape UNet}
& 89.76 & 89.87 & 90.15 & 90.06
& 89.04 & 89.19 & 89.47 & 89.38 \\

\multicolumn{1}{l}{\itshape FR-UNet}
& 90.14 & 90.04 & 90.37 & 90.29
& 89.57 & 89.47 & 89.80 & 89.69  \\

\multicolumn{1}{l}{\itshape MA-Net}
& 89.41 & 89.88 & 89.97 & 90.05
& 88.76 & 89.23 & 89.29 & 89.36  \\

\multicolumn{1}{l}{\itshape SA-UNet}
& 85.90 & 85.57 & 86.55 & 85.76
& 84.91 & 84.64 & 85.64 & 84.83  \\

\multicolumn{1}{l}{\itshape W-Net}
& 84.42 & 85.62 & 85.65 & 85.67
& 83.50 & 84.72 & 84.74 & 84.75  \\
\bottomrule
\end{tabular*}
\vskip -0.1in
\label{tab:model_results_various_loss}
\end{table}

\subsection{Robustness of different architectures to domain shifts}

We further investigated the generalization capabilities of the trained models on additional datasets that were not part of the training procedure. Due to the domain shift, we expect that a model trained on one dataset would not achieve the same level of performance when tested on another dataset. For this purpose, we additionally used the much smaller CHASE DB and the DRIVE datasets (see Methods), which look clearly different  from the FIVES dataset and thus should induce strong domain shifts (Fig. \ref{fig:dataset_masks}). Trained models reached a similar performance as reported in the literature (Table \ref{tab:in-domain_datasets}).  For this analysis, we focused on the DiceBCE loss function for all models for simplicity. We trained all models individually on each dataset and evaluated their performance on the two remaining datasets.

\begin{figure}[htbp]
    \centering
    \includegraphics[width=\textwidth]{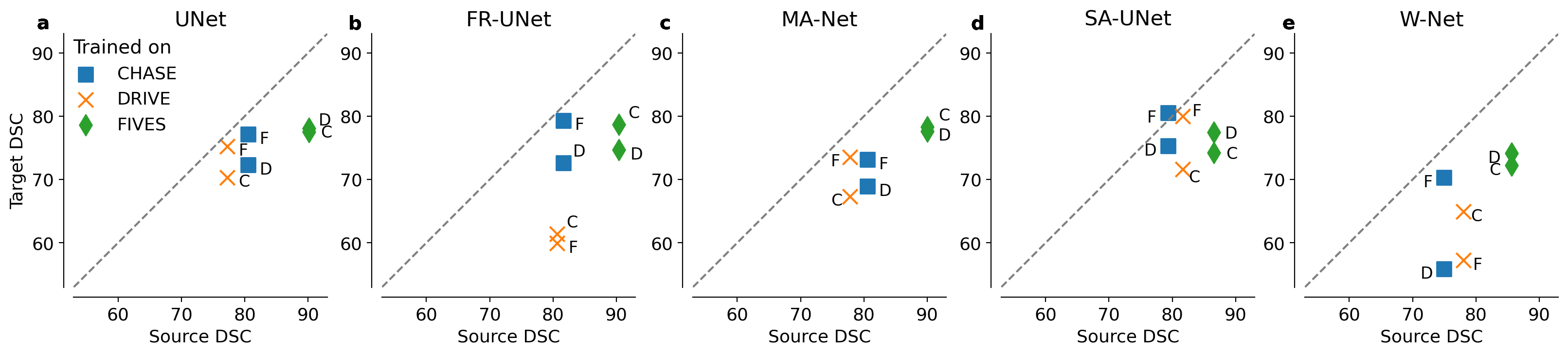}
    \caption{Cross dataset generalization in terms of Dice for each model. Each marker represents a tuple of the source (training) and the target (testing) dataset. The former is indicated by the marker type and the latter is indicated with a subscript letter. Hence the blue square with subscript F shows the mean Dice over the FIVES dataset for a model that was trained on CHASE DB. The vertical distance to the diagonal quantifies the domain gap.}
    \label{fig:cross-data}
\end{figure}

\begin{table}
    \begin{minipage}{0.5\textwidth}
        \vspace{0pt}
        \centering
        \caption{Dice of the In-domain performance for all datasets.}
        \begin{tabular}{cccc}
        \toprule
            Model & FIVES & DRIVE & CHASEDB1\\
        \midrule
            UNet & 90.15 & 77.23 & 80.55\\
            FR-UNet & 90.37 & 80.62 & 81.65\\
            MA-Net & 89.97 & 77.78 & 80.56\\
            SA-UNet & 86.55 & 81.65 & 79.30\\
            W-Net & 85.65 & 77.98 & 74.98\\
        \bottomrule
        \end{tabular}
        
\label{tab:in-domain_datasets}
    \end{minipage}
        \hfill
    \begin{minipage}{0.5\textwidth}
            \vspace{11.5pt}
            \centering
            \caption{Number of high quality images per category in the train split of FIVES (n=600).}
            \begin{tabular}{ccccc}
            \toprule
                 & Illumination/ Color & Blur & Contrast \\
            \midrule
                 AMD & 142 & 134 & 150 \\
                 DR & 127 & 117 & 149 \\
                 Glaucoma & 82 & 109 & 131 \\
                 Normal & 145 & 150 & 150 \\
            \bottomrule
                $\Sigma$ & 496 & 510 & 580 
            \end{tabular}
            
\label{tab:img_quality}
    \end{minipage}

\end{table}

 
 We found that the simple UNet and the SA-UNet handled dataset shifts better than other models, as their cross-dataset performance was closer to their in-domain performance (Fig.~\ref{fig:cross-data}). Interestingly, the SA-UNet model, which showed comparatively low performance in the in-domain settings, did not have much of a performance gap in the cross-dataset setting. We found that the W-Net and the FR-UNet were the most sensitive to dataset shifts. Models trained on the FIVES datasets (green diamonds) generalized well to both the DRIVE and CHASEDB1 datasets (Fig.~\ref{fig:cross-data}). This is likely due to the large number of samples available for training and their high quality annotation. Interestingly, for the Unet, the cross-dataset performance of a model trained on FIVES, when evaluated on DRIVE and CHASE-DB, was no worse than the in-domain performance of a model trained directly on these datasets  (Fig.~\ref{fig:cross-data}a). This suggests that generalization capabilities across datasets are an important factor to be evaluated when developing new retinal vessel segmentation methods.

\subsection{Robustness of different architectures to disease-related domain shifts}

Next, we evaluated the performance of all models in four subgroups related to ophthalmological conditions (AMD, DR, glaucoma and healthy images). 
We considered two setups: First, we trained each model on three subgroups and tested it on the remaining subgroup (denoted as 3 vs. 1) to mimic the scenario where a model is applied to fundus images with diseases not present in the training data. In addition, we trained each model on the whole training set and evaluated its performance for the individual subgroups in the test set.

\begin{figure}[htbp]
    \centering
    \includegraphics[scale=0.6]{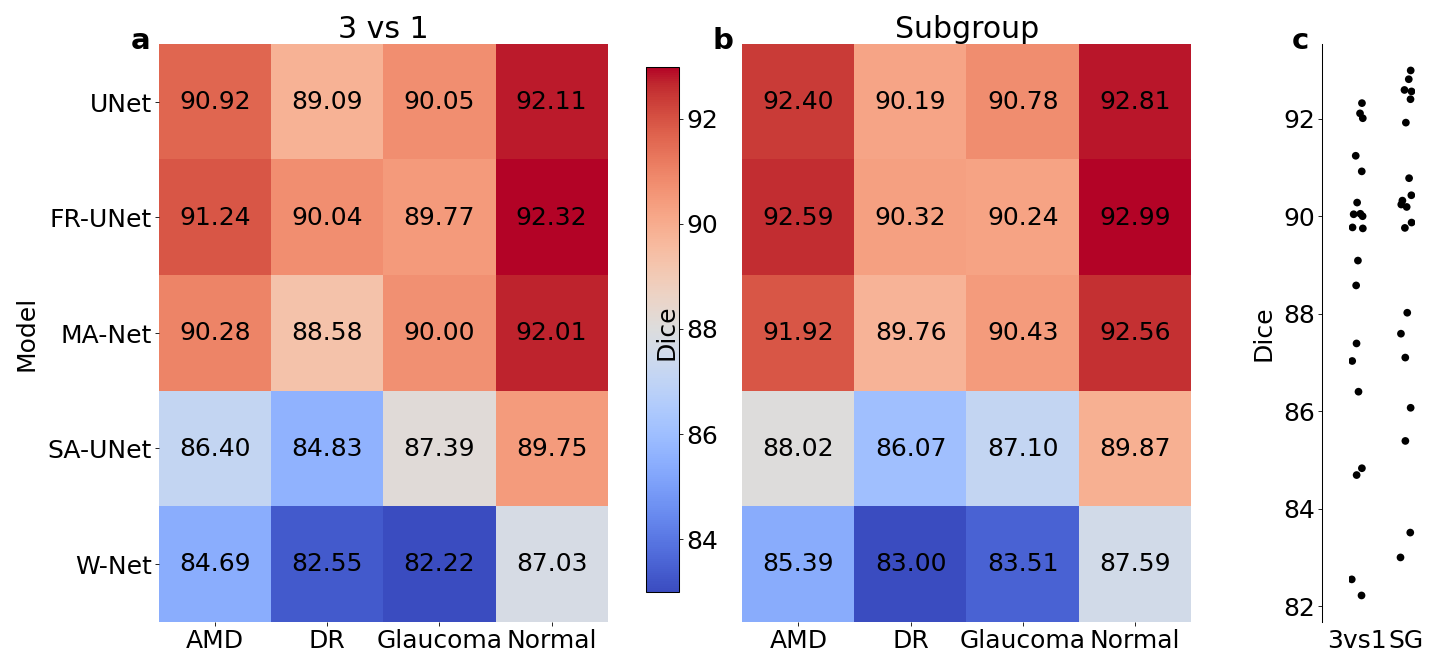}
    \caption{Generalization across diseases. \textbf{a)} The segmentation performance in the 3 vs. 1 scenario, when models were trained on three pathological conditions only and tested on the remaining one. \textbf{b)} The average performance within the subgroup after the regular training procedure.    
    \textbf{c)} Contrasts both settings directly, where each pair of points corresponds to entries in the heatmaps in \textbf{a)} and \textbf{b)}.}
    \label{fig:sub-group-disease}
\end{figure}

We found that the median performance in the 3 vs. 1 setting was similar to the in-domain performance of the respective model, with W-Net having the maximum performance drop of 4.03\% (Fig.~\ref{fig:sub-group-disease}~\textbf{a, b, c}). Thus, domain shifts with respect to disease did not affect the segmentation quality substantially, at least when the model was trained on a sufficiently large dataset. 

We also found that all models were better at segmenting fundus images from healthy eyes and eyes with AMD compared to fundus images from DR and glaucoma patients (Fig.~\ref{fig:sub-group-disease}~\textbf{a}, \textbf{b}). While UNet, FR-UNet and MA-Net performed best also in the subgroup analysis, SA-UNet and W-Net perform worse in the subgroup setting, following the same worse performance as in the standard setting (Table~\ref{tab:in-domain_datasets}).

\subsubsection*{Robustness of different architectures to image quality}

\begin{figure}[htbp]
    \centering
    \includegraphics[width=\textwidth]{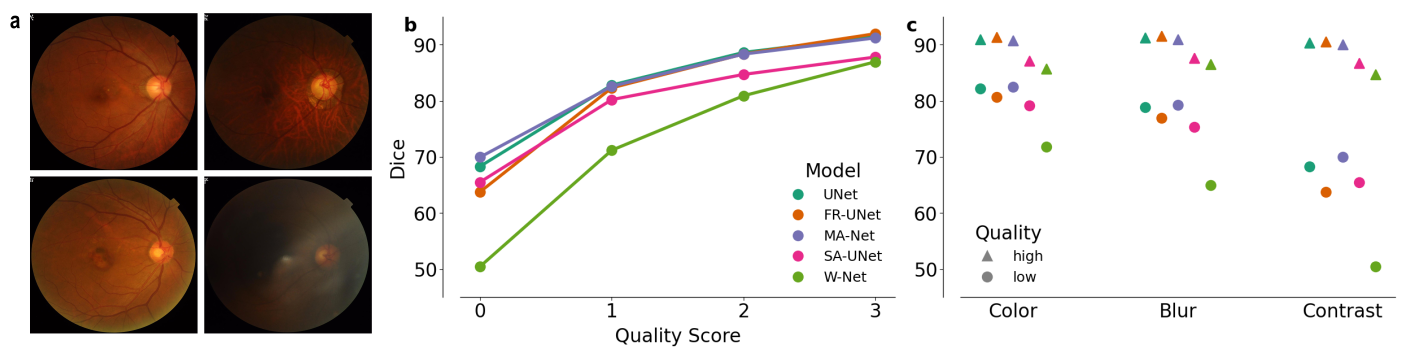}
    \caption{Impact of image quality on segmentation performance. \textbf{a}: Example images with varying quality. \mbox{Column 1:} High quality in all aspects, Column 2: Low Illumination and Color (first image), low quality in all three categories (second image). \textbf{b}: Overall image quality, \textbf{c}: Image quality split up into its three different aspects Illumination and Colour, Blur and Contrast.}
    \label{fig:image-quality}
\end{figure}

Finally, we evaluated how robustly the different models performed in images of different quality. Each image in the FIVES dataset is labeled as ``high'' or ``low'' quality with respect to illumination and color, blur, and low contrast, scored by an automatic algorithm \citep{7349228} (Fig. \ref{fig:image-quality}c, Table \ref{tab:img_quality}). We combined all image quality metrics into an overall quality score, which a score of $0$ representing low quality in all quality criteria, and a score of $3$ representing good quality in all criteria. 

We evaluated the segmentation performance of all models for each image quality level  (Figure~\ref{fig:image-quality}~\textbf{a}). Not surprisingly, the models performed worst for the poorest quality images. All models benefited to a similar degree from increasing  image quality, with the exception of the W-Net, which appeared to be most susceptible to image quality degradation and benefited most from quality improvements. Interestingly, the FR-UNet, being one of the overall top-performing models, performed worse than the generally low-performing SA-UNet model for the subgroup with the poorest overall image quality. This may be due to the fact that the FR-UNet operates at full resolution, potentially making  the model's intermediate representations more susceptible to noise, as it is averaged out when the resolution is  downsampled in the other models.

We then investigated the three individual components of image quality separately. Poor contrast affected all models strongly and led to the highest drop in segmentation performance (Figure~\ref{fig:image-quality}~\textbf{b}), while the effect of blur and illumination was less pronounced. Nevertheless, all components of image quality had large effects on segmentation performance (much larger than the effect of different diseases), underscoring the importance of good image quality for tasks such as vessel segmentation.

\section{Discussion}

In this paper, we comprehensively compared the state of the art in retinal vessel segmentation \citep{FR-Unet, saunet, little-wnet} and investigated commonly used model architectures and loss functions in three publicly available datasets \citep{drive_data, fives, chase_data}. 
Most existing evaluation papers \citep{hegde2023systematic} primarily focus on summarizing available research and generally conclude by endorsing deep learning (DL)-based approaches, such that a rigorous benchmark of a variety of DL methods and their robustness has been much needed. We evaluated the models' in-domain performance as well as their generalization to different datasets as well as their robustness to images with different diseases and different levels of image quality.

We found that the choice of loss function did not crucially affect the segmentation quality, but the clDice and DiceBCE loss marginally outperformed the other optimization criteria. Similarly, we could not identify a single optimal model architecture, with UNet, FR-UNet and MA-Net performing similarly well. This implies that, interestingly, the original UNet without modification remains state-of-the-art for the vessel segmentation. 

Deep learning-based segmentation models are known to be susceptible to dataset shifts \citep{Boone2023-qh, Koch2024-hd}, and, as we show in this study, retinal vessel segmentation models are not an exception. Interestingly, while variations in image quality due to differences in the imaging setup have the most impact on the segmentation output, variations due to disease manifestation/prevalence do not significantly affect the model's performance. Therefore, high image quality is crucial for a successful vessel segmentation.
Among the publicly available datasets training on FIVES, which has the largest sample size at a high image quality, allowed for the best generalization to other datasets, when averaged over all models. Hence, we recommend to choose FIVES as a training set for studies where cross dataset generalization matters. 

While models do not generalize well from high to poor image quality, they generalize quite well to unseen diseases. When curating data for practice this means that ensuring high image quality should be the primary concern. Generalization to diseases not  present in the training data can likely be expected. Foundation models for medical image segmentation trained on wide range of imaging data \citep{Ma2024-jo} may further improve generalization of retinal image segmentation, but currently face challenges accurately segmenting retinal vessels even with fine-tuning \citep{Shi_2023}. 

Finally, future studies should consider how different segmentation algorithms affect downstream tasks and evaluate them directly as part of the clinical pipelines. For example, probabilistic segmentation approaches have been evaluated as part of the eye disc and cup for glaucoma diagnosis \citep{wundram2024leveraging}. Similar evaluation pipelines for vessel segmentations would be needed to directly address the clinical potential of different segmentation models.

\section*{Acknowledgements}
This work was supported by a grant from the Carnegie Corporation of New York (provided through the African Institute for Mathematical Sciences), the German Science Foundation (BE5601/8-1 and the Excellence Cluster 2064 ``Machine Learning --- New Perspectives for Science'', project number 390727645), the Carl Zeiss Foundation (``Certification and Foundations of Safe Machine Learning Systems in Healthcare'') and the Hertie Foundation. PB is also a member of EKFS-Kolleg ``ClinBrain''.

\clearpage

\bibliographystyle{plainnat}
\bibliography{references.bib}


\clearpage

\end{document}